# Non-thermal response of YBa$_2$Cu$_3$O$_{7-\delta}$ thin films to picosecond THz pulses

P. Probst[1], A. Semenov[2], M. Ries[3], A. Hoehl[4], P. Rieger[5], A. Scheuring[1], V. Judin[6], S. Wünsch[1], K. Il'in[1], N. Smale[5], Y.-L. Mathis[5], R. Müller[4], G. Ulm[4], G. Wüstefeld[3], H.-W. Hübers[2, 7], J. Hänisch[8], B. Holzapfel[8], M. Siegel[1] and A.-S. Müller[6]

[1] *Institut für Mikro- und Nanoelektronische Systeme, Karlsruher Institut für Technologie (KIT), Hertzstraße 16, 76187 Karlsruhe, Germany*
[2] *DLR e.V. (German Aerospace Center), Institut für Planetenforschung, Rutherfordstraße 2, 12489 Berlin-Adlershof, Germany*
[3] *Helmholtz-Zentrum Berlin (HZB), Albert-Einstein-Straße 15, 12489 Berlin, Germany*
[4] *Physikalisch-Technische Bundesanstalt (PTB) Berlin, Abbestraße 2-12, 10587 Berlin, Germany*
[5] *Institut für Synchrotronstrahlung (ISS), Karlsruher Institut für Technologie (KIT), Hermann-von-Helmholtz-Platz 1, 76344 Eggenstein-Leopoldshafen, Germany*
[6] *Laboratorium für Applikationen der Synchrotronstrahlung, Karlsruher Institut für Technologie (KIT) (KIT), Engesser Straße 15, 76131 Karlsruhe, Germany*
[7] *Technische Universität Berlin, Institut für Optik und Atomare Physik, Hardenbergstrasse 36, 10623 Berlin, Germany*
[8] *Leibniz-Institut für Festkörper- und Werkstoffforschung Dresden, IFW Dresden, Helmholtzstraße 20, 01069 Dresden, Germany*

The photoresponse of YBa$_2$Cu$_3$O$_{7-\delta}$ thin film microbridges with thicknesses between 15 and 50 nm was studied in the optical and terahertz frequency range. The voltage transients in response to short radiation pulses were recorded in real time with a resolution of a few tens of picoseconds. The bridges were excited by either femtosecond pulses at a wavelength of 0.8 µm or broadband (0.1 – 1.5 THz) picosecond pulses of coherent synchrotron radiation. The transients in response to optical radiation are qualitatively well explained in the framework of the two-temperature model with a fast component in the picosecond range and a bolometric nanosecond component whose decay time depends on the film thickness. The transients in the THz regime showed no bolometric component and had amplitudes up to three orders of magnitude larger than the two-temperature model predicts. Additionally THz-field dependent transients in the absence of DC bias were observed. We attribute the response in the THz regime to a rearrangement of vortices caused by high-frequency currents.

## I. INTRODUCTION

The photoresponse of superconducting YBa$_2$Cu$_3$O$_{7-\delta}$ (YBCO) thin films was already extensively studied in the spectral range from optical to infrared wavelength [1-4]. Typically, films were structured to form micrometer-large bridges with thicknesses between 40 and 100 nm. In all these experiments the photon energy of incoming radiation was above the superconducting energy gap in YBCO and hence radiation was directly absorbed in the microbridges by electrons. The photoresponse in this wavelength range is well understood. It was described in the framework of the two-temperature model (2T-model) first discussed by Perrin and Vanneste[5] for sinusoidal perturbations and by Semenov et al.[3] for optical pulse excitations. The general idea is that electrons excited by absorbed radiation destroy cooper pairs and produce hot electrons in the microbridge. An increased quasi-equilibrium electron temperature $T_e$ is established within the thermalization time $\tau_{th}$. The excess energy is then transferred to the phonons within the electron-phonon interaction time $\tau_{e-ph}$ that increases the quasi-equilibrium phonon temperature $T_{ph}$ (hence 2T-model). In YBCO these two interaction times are in the sub-picosecond ($\tau_{th}$ = 0.56 ps) and picosecond ($\tau_{e-ph}$) range as it was measured by electro-optical sampling technique[6]. Finally, within the phonon escape time $\tau_{esc}$ the energy is released to the substrate which permanently remains at the bath temperature. Because all three processes occur simultaneously, the time for the microbridge to return to global equilibrium is a complicated function of aforementioned particular times. In YBCO the specific heat capacity of the phonons $C_{ph}$ is much larger than the electron specific heat capacity $C_e$ ($C_{ph}/C_e$ = 38 (Ref. 6)). Therefore, at the time scale of $\tau_{e-ph}$ the electron and phonon systems are thermally decoupled and the energy backflow from the phonon to the electron system can be neglected. This explains the fast picosecond component in the photoresponse of YBCO microbridges. The slow component is largely due to the cooling of phonons via escape to the substrate. Responses described by the detection mechanism which includes creation of hot electrons and subsequent electron-phonon cooling was reported for wavelengths up to $\lambda = 20\,\mu m$ ($f = 15$ THz) (Ref. 7).

Although slight differences in the response of a YBCO microbridge to optical and infrared radiation pulses as well as non-bolometric features of the response to THz radiation have been noted in earlier publications[8,9], it was generally accepted that the detection mechanism remains unchanged even for sub-millimeter wavelenghts (sub-terahertz frequency range). The THz radiation pulses available at that time were not sufficiently short to allow for quantitative evaluation of the non-bolometric features and to find out to what extent they may affect the response time of the microbridge. First the discovery of intensive coherent synchrotron radiation (CSR) in the form of picosecond pulses[10] has made it possible to analyze the response time of a YBCO microbridge to sub-gap excitation[11]. The measurements reported in Ref. 11 were performed at two different synchrotron radiation facilities MLS[12] and ANKA[13]. They all showed that, in contrast to the optical excitation, the response of a YBCO microbridge to CSR pulses does not have a bolometric component. This observation motivates further work to get insight in the detection mechanism at THz waves and to promote further optimization of fast CSR detectors.

In this paper we report on the comparison study of the photoresponse of the very same YBCO microbridges to pulsed over-gap (optical frequency range) and sub-gap (THz frequency range) excitation at the same pulse energies. We show that the response to optical radiation is qualitatively well explained by the 2T-model and that this model completely fails in describing the response to THz radiation. We further demonstrate that microbridges detect picosecond CSR pulses in the absence of any DC bias current and that the electric THz-field rather than the pulse energy controls the response to CSR. We suggest direct interaction of vortices with the high-frequency electric field to explain photoresponse of YBCO microbridges to picosecond CSR pulses.

## II. YBCO THIN FILMS ON SAPPHIRE

Thin YBCO films were fabricated by the pulsed-laser deposition (PLD) technique using a KrF excimer laser (wavelength $\lambda = 248$ nm) with a pulse energy density of $\approx 1\,J/cm^2$ on the target surface. The PLD system is equipped with a carousel of six targets allowing *in-situ* fabrication of multi-layer samples. The heated substrate holder is situated about 50 mm from the target in on-axis position. To ensure a good thermal contact during deposition the substrate was mounted on the heater with silver paste. The vacuum chamber was evacuated by a turbo pump to a base pressure below $1 \cdot 10^{-5}$ mbar.

In spite of good crystalline matching of $SrTiO_3$ or $LaAlO_3$ to YBCO (mismatch less than 1%)[14], we chose both-side polished R-plane sapphire as substrate for our YBCO thin films with a strong lattice mismatch of 12% (Ref. 14). This is due to the low dielectric losses of sapphire ($\varepsilon_r = 10.06$, $\tan\delta = 8.4 \cdot 10^{-6}$ at 77 K)[15] which are essential for the back-illumination of our YBCO detectors at THz frequencies. However, due to the crystalline mismatch between sapphire and YBCO and diffusion of aluminum into the YBCO film at high deposition temperatures[14], it is not possible to grow high-quality superconducting YBCO thin films directly on sapphire. Buffer layers to improve the matching of the crystalline structure have to be used in the deposition process.

Therefore, a $CeO_2$ buffer layer with a thickness of 8 nm was deposited at a substrate temperature of 800°C and an oxygen pressure of $p_{O2} = 0.9$ mbar. To further improve the crystalline matching with the YBCO film, an additional buffer layer made of a $d_1 = 25$ nm thick $PrBa_2Cu_3O_{7-\delta}$ (PBCO) layer was deposited on top of the $CeO_2$ layer. For this, the deposition temperature was kept constant and the partial oxygen pressure was reduced to $p_{O2} = 0.7$ mbar. A YBCO film with a thickness $d$ between 10 and 80 nm was deposited on top of the PBCO buffer layer at the same temperature and oxygen pressure. For protection of the YBCO thin film during patterning a second passivating PBCO layer of a thickness $d_2 = 25$ nm was deposited on top of the YBCO film. After the second PBCO layer deposition the oxygen pressure was increased to 900 mbar and the substrate temperature was ramped down to 550 °C with a rate of 10°C/min. The temperature was kept constant at 550 °C for ten minutes for annealing of the obtained multi-layer structure. Afterwards, the heater was ramped down to 400 °C before switching off and cooling down exponentially to room temperature. The vacuum chamber was then pumped down to pressures about $5 \cdot 10^{-5}$ mbar and a 140 nm thick Au layer was grown *in-situ* using the same PLD technique.

The as-deposited films were characterized using a quasi four-probe measurement configuration to determine the dependence of the film resistance on temperature. The critical temperature $T_c$ was defined as the temperature at which the resistance reaches 1% of the normal state resistance right above the transition. The transition width was determined as the temperature interval between 90% and 10% of the normal state resistance. The dependence of the critical temperature $T_c$ (squares) and the transition width (circles) on the YBCO film thickness is shown in fig. 1. For YBCO film thicknesses above 30 nm the critical temperature is nearly constant and well above the liquid nitrogen temperature of 77 K reaching $T_c$ of 87 K. The critical temperature decreases for films thinner than 30 nm. This can be explained by the above mentioned lattice mismatch leading to oxygen deficiency in the YBCO film, which results in a reduction of $T_c$ (Ref. 8). However, due to the introduction of the $CeO_2$ and PBCO buffer layers the lattice mismatch between YBCO and the buffer layers could be significantly reduced allowing us to reach critical temperatures $T_c$ = 79 K even for the thinnest film with the thickness $d$ = 10 nm that corresponds to only 8 unit cells.

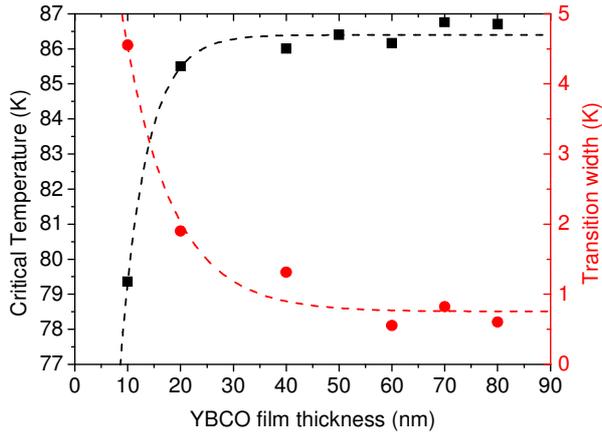

FIG. 1. (Color online) Dependence of zero resistance critical temperature (squares) and transition width (circles) on YBCO film thickness for as-deposited films. The dashed lines are to guide the eye.

## III. THE YBCO SAMPLES

For the photoresponse measurements in the optical frequency range the radiation was directly focused on the microbridge since the wavelengths are smaller than the lateral dimensions of the microbridges. However, for the photoresponse studies in the THz range, owing to much larger wavelength, a planar antenna is required to couple in the radiation.

### A. Sample layout and THz antenna simulation

To provide broadband coupling in the THz frequency range we implemented a planar log-spiral antenna. The antenna was embedded into a coplanar readout line. A schematic layout of the antenna structure is shown in fig. 2a. The antenna was designed with CST Microwave Studio®[16] to couple radiation in the frequency range from 0.1 to 2 THz that encompasses the spectrum of a typical CSR source[12, 13]. Fig. 3a shows the simulated impedance for the antenna on semi-infinite sapphire substrate. The real part is almost constant ($R_a \approx 65\ \Omega$) over the entire frequency range. This agrees well with earlier published data[17]. The imaginary part is nearly zero. Fig. 3b shows the simulated reflection parameter ($|S_{11}|$) for a detector with a frequency-independent real impedance of 50 Ω. From 0.15 THz up to more than 2.5 THz the reflection parameter is well below -10 dB that corresponds to a coupling efficiency higher than 90% in this frequency range.

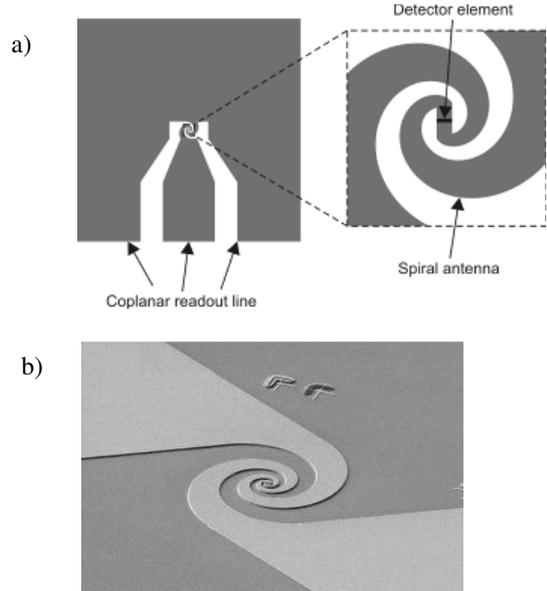

FIG. 2. (a) Left panel - sample layout. Gray color shows Au metallization. The log-spiral antenna is embedded into a coplanar readout line. Right panel - zoom in of the antenna with the microbridge (black) at the center feed point.
(b) SEM image of the logarithmic-spiral antenna used for the THz pulse measurements. The bright part is the Au metallization while the darker part shows the substrate.

### B. Preparation and characterization of the YBCO samples

The as-deposited multilayers (for details see section II) were patterned by several electron-beam lithography and etching steps. For the measurements discussed in section V YBCO films of 15, 30 and 50 nm in thickness were used. The active/detecting area of the YBCO sample (microbridge) was opened via removing the Au layer by wet etching with a $I_2$-KI solution. Ion milling was used to define the antenna and the coplanar readout. A SEM image of the center part of the sample with logarithmic spiral antenna is shown in fig. 2b. To reduce oxygen loss in the YBCO layer during the ion milling we actively cooled the substrate. The superconducting microbridges were 2 μm long rectangles with a width of $w$ = 4.5 μm and had a normal state resistance $R_n$ just above the superconducting transition in the range from 100 to 200 Ω (see table 1).

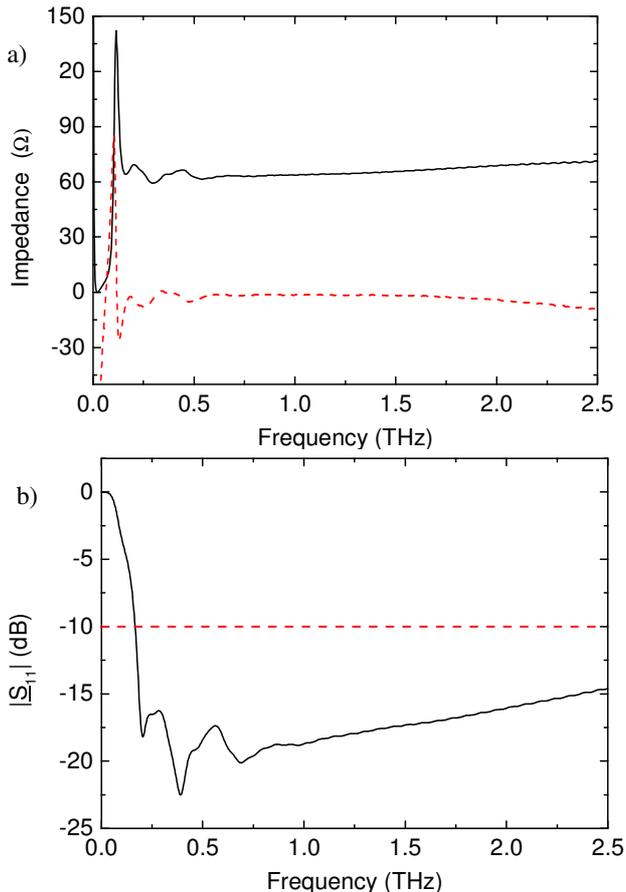

FIG. 3. (Color online) Simulated parameters of the log-spiral antenna which was used for the THz measurements. (a) Real (solid line) and imaginary part (dashed line) of the impedance; (b) |$\underline{S}_{11}$| parameter for the detector impedance of 50 Ω (solid line). The dashed line shows the -10 dB level.

TABLE 1: Characteristics of the YBCO thin film samples used for the photoresponse experiments

| YBCO film thickness (nm) | Critical temperature (K) | Critical current at 77 K (mA) | Critical current density at 77 K (MA/cm²) | Normal state resistance (Ω) |
|---|---|---|---|---|
| 15 | 83.4 | 1.6 | 2.4 | 140 |
| 30 | 85.2 | 3.2 | 2.3 | 170 |
| 50 | 86.4 | 6.9 | 2.9 | 100 |

## IV. EXPERIMENTAL SETUP

The sapphire substrates with antenna-coupled microbridges were glued to the rear flat side of an elliptical silicon lens and mounted in a shielded holder to reduce interference with the electromagnetic fields at the synchrotron. Optical radiation was directed to the microbridge through a small pinhole in the holder. The measurements were performed in a continuous flow cryostat with optical windows from fused quartz. The temperature was controlled by a temperature sensor which was built in the holder. The bias was arranged via a room-temperature bias-tee with a bandwidth of more than 25 GHz. The detector signal was amplified with a room temperature amplifier (bandwidth 0.1 - 8 GHz) and fed into a sampling oscilloscope with a 20 GHz bandwidth (LeCroy SDA 100G). The electronic components define the effective bandwidth of the readout[3]

$$f = \left[\sum_i f_i^{-2}\right]^{-1/2} = 6.4\,\text{GHz} \qquad (1)$$

where $f_i$ are their bandwidths. Accordingly, the effective time resolution of the readout system amounted to ≈ 80 ps.

We systematically studied the response of our microbridges to broadband picosecond THz pulses and to femtosecond optical pulses while preserving either the same pulse energy or the same bias current and temperature in both cases.

The response to CSR pulses was measured at two synchrotron radiation facilities ANKA[13] and MLS[12, 18, 19]. ANKA, the synchrotron light source of the Karlsruhe Institute of Technology can be operated in the beam energy range between 0.5 and 2.5 GeV. The ANKA storage ring can accommodate up to 184 bunches separated by 2 ns corresponding to the 500 MHz of the RF system. The Metrology Light Source (MLS) of the Physikalisch-Technische Bundesanstalt is the first electron storage ring worldwide optimized to the low-alpha operation mode and, hence, to the production of high power THz CSR. The storage ring can accommodate any number of bunches of up to 80 spaced by 2 ns.

To excite the microbridges in the optical range we used a TiAl₂O₃ femtosecond pulsed laser which delivered 26 fs pulses with a repetition rate of 80 MHz at a wavelength of 0.8 μm.

In both cases the pulse energy was precisely controlled by measuring the mean radiation power in front of the cryostat and accounting for all optical losses on the way to the microbridge.

### A. Coupled pulse energy in the optical frequency range

With the maximum averaged input power of 23 mW and the 80 μm beam diameter, we estimated the pulse energy of $E = 0.69$ pJ to be delivered to the microbridge area of $9 \cdot 10^{-12}$ m² (width: 4.5 μm, length: 2 μm). We assumed a Gaussian beam profile and took into account losses at the cryostat window. To obtain the absorbed energy we used the absorption coefficient for films with the skin depth much less than the wavelength[20]

$$A = \frac{4Z_0 R_{sq}}{(Z_0 + (n+1)R_{sq})^2}, \quad (2)$$

where $Z_0$ is the impedance of free space, $R_{sq}$ the square resistance of the microbridge in the normal state and $n$ the refraction index of sapphire. With $Z_0 = 120\pi\,\Omega$, $n = 1.7$ and $R_{sq} = 320\,\Omega$ we found the absorption coefficient $A \approx 0.3$. The output power of the laser was varied with a set of gray filters resulting in absorbed energies per pulse from 206 to 9 fJ.

### B. Coupled pulse energy in the THz frequency range

To compute the THz pulse energy, which was delivered to the microbridge, we took into account the mode-mismatch between the CSR beam and the directional pattern of the elliptical lens as well as the reflections at the cryostat window and at the lens surface. The THz synchrotron radiation has a repetition rate of 500 MHz. The maximum averaged output power was 3.6 mW resulting in a pulse energy of 7.2 pJ. By decreasing the electron current in the storage ring, the output power was reduced to a detectable minimum of 30 μW (60 fJ pulse energy).

The coupling between two Gaussian modes with different beam waists is[21]

$$K = \frac{4 w_1^2 w_2^2}{(w_1^2 + w_2^2)^2}. \quad (3)$$

In our case the CSR beam waist is $w_1 = 4.8$ mm while the equivalent waist of the lense pattern at 1 THz equals $w_2 = 2.2$ mm (Ref. 22). This results in the coupling factor $K = 0.57$. The reflectivity of the window and the lens surface was calculated as

$$\Gamma = \left(\frac{\sqrt{\varepsilon_r}-1}{\sqrt{\varepsilon_r}+1}\right)^2 \quad (4)$$

with the corresponding dielectric constant $\varepsilon_r$ resulting in a total reflection loss of $\Gamma_{tot} = 0.37$.

Given the frequency independent impedance of the microbridge $Z_d$, the coupling efficiency of the antenna can be computed with standard antenna theory[23] as

$$\eta = \frac{\mathrm{Re}(Z_d)}{[(\mathrm{Re}(Z_a)+\mathrm{Re}(Z_d))^2 + (\mathrm{Im}(Z_a)+\mathrm{Im}(Z_d))^2]^{1/2}} \quad (5)$$

where $Z_a$ is the impedance of the antenna. While for over-gap excitation $Z_d = \mathrm{Re}(Z_d) + i\,\mathrm{Im}(Z_d)$ with $\mathrm{Re}(Z_d) \approx \mathrm{Im}(Z_d) \approx R_n$ (Ref. 17), in our case of sub-gap excitation the impedance of the microbridge should have much smaller real component. Its value is defined by the nature of the resistive state. To avoid uncertainties resulting from poor knowledge of the impedance in the resistive state, we will use the impedance for over-gap excitation. Taking into account the almost real impedance of the antenna $\mathrm{Re}(Z_a) = 65\,\Omega$ in the range of interest, the coupling efficiency results in $\eta \approx 0.5$ for the 30 nm thick sample. We use equations (3)-(5) to obtain the total coupling factor $K(1-\Gamma_{tot})\eta \approx 0.2$. This results in the coupled energies between 1.4 pJ and 12 fJ. Thus evaluated coupled energy is the upper boundary for the sub-gap case; we will use it to present our experimental results in the next section.

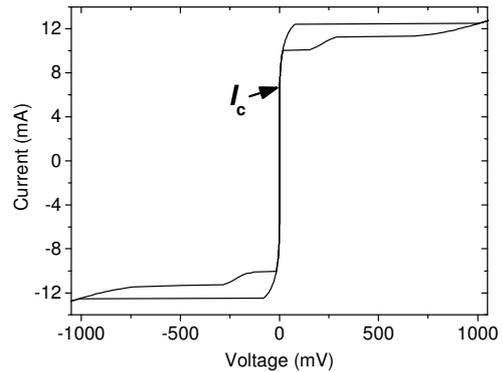

FIG. 4. Current-voltage characteristic of the 30 nm thick sample acquired at 72 K with current bias.

## V. EXPERIMENTAL RESULTS

In order to obtain experimental evidence of the non-bolometric nature of the THz photoresponse we perform a thorough comparison of the response transients to short THz (sub-gap) and visible-light (overgap) excitations. The transients at these two spectral ranges were acquired with the very same microbridge at the same operation conditions. Moreover, the energies of the pulses had comparable values. In the following, we describe in detail pronounced differences between the response transients at these two spectral ranges.

### A. Absence of the slow component in the THz regime and transient magnitudes

We have found that in the THz frequency range the time evolution of the response transient does not vary with the pulse energy and has an energy independent duration of 80 ps (full width at half maximum (FWHM)), see fig. 5Ia) and Ib). In the optical range the transient changes with increasing pulse energy. At energies less than 75 fJ only the fast component with a FWHM of 80 ps is observed (see fig. 5 IIa)). At larger energies the bolometric component with a decay time of a few nanoseconds appears (see fig. 5 IIb)). Another significant difference between the two regimes is the transient magnitude. At the same operation conditions and for similar pulse energies the transient magnitudes in the THz frequency range are up to an order of magnitude larger than in the optical range (see fig. 5).

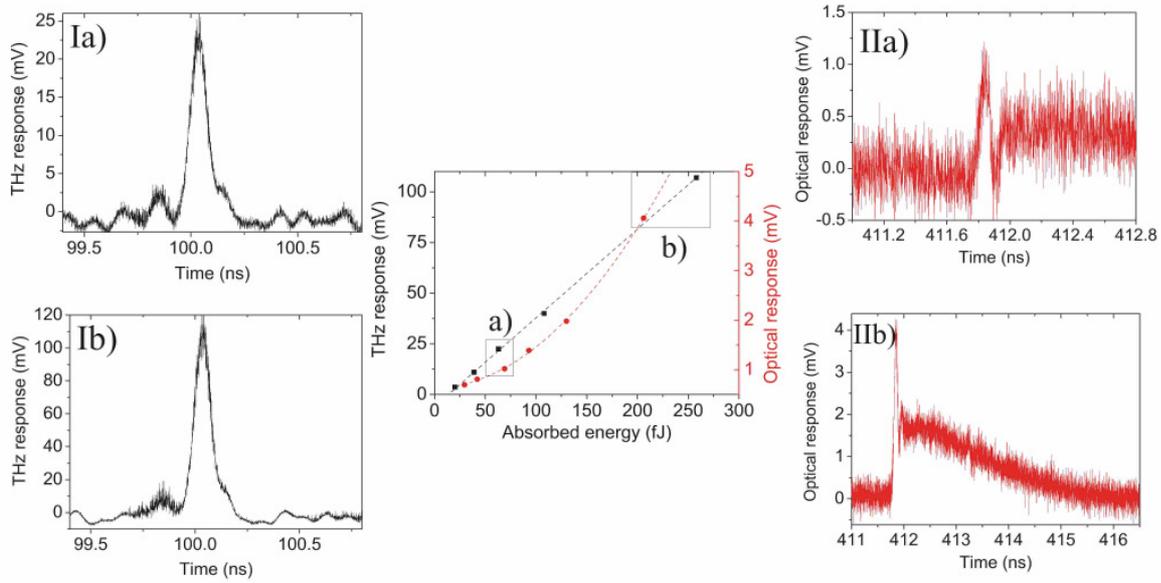

FIG. 5. (Color online) Response transients recorded with the 30 nm sample at THz (I) and optical (II) frequencies for different pulse energies indicated in the centric graph. The transients were acquired at $T = 0.85T_c = 72$ K and at a bias current of 12 mA (THz regime) and 7 mA (optical wavelengths). The central plot shows the pulse-energy dependencies of the transient magnitudes in these two regimes. The dashed lines emphasis the linear (THz) and non-linear (optical) dependencies.

Not only the magnitudes differ, but also the pulse-energy dependences of the transient magnitudes are different. In the THz-regime the transient magnitude scales linearly with the pulse energy while in the optical range the dependence is superlinear. The central plot in fig. 5 illustrates these peculiarities for the 30 nm thick sample at $T = 72$ K. The IV-characteristic, which was acquired at this temperature, is displayed in fig. 4.

Also the dependence of the transient magnitude on the bias current shows a deviation from the typical superlinear dependence in the optical frequency range. In fig. 6 the superlinear dependence of the optical response (circles) on the bias current is displayed. Even for the largest absorbed energy levels in the THz frequency range (1.4 pJ) this superlinear behaviour could not be reproduced in the THz-regime where the transient magnitude increases linearly with high bias currents.

### B. Zero-bias response to THz radiation

Zero-bias conditions have been realized by disconnecting the bias source and leaving the bias-line open. As expected for an isotropic bolometer where no thermoelectric voltage is present, we found no response transient to optical-light pulses without bias even for the largest available pulse energy. Contrary, there was a clear response to THz-pulses with transient magnitudes in the range of a few millivolts (see fig. 7). The sign of the transient changed when we introduced a $\pi$-shift in the phase of the THz electric field by substituting one of the flat metallic mirrors (p-polarization) in the optical path with a dielectric mirror from strontium titanate (STO). Fig. 7 shows the transients obtained with these two mirrors for the pulse energy of 1.4 pJ.

## VI. DISCUSSION

To facilitate the discussion we summarize the major differences which appear in the response of YBCO microbridges to short THz pulses as compared to optical excitation. (i) There is no bolometric component in the THz response. The shape of the response transient does not depend on variations of either pulse energy or bias current. (ii) As function of both the absorbed energy and the applied bias current, the superlinear dependences of the transient magnitude in the optical range change to linear dependences in the THz range. (iii) There is a response at zero bias in the THz regime while no response transients have been observed in the optical range.

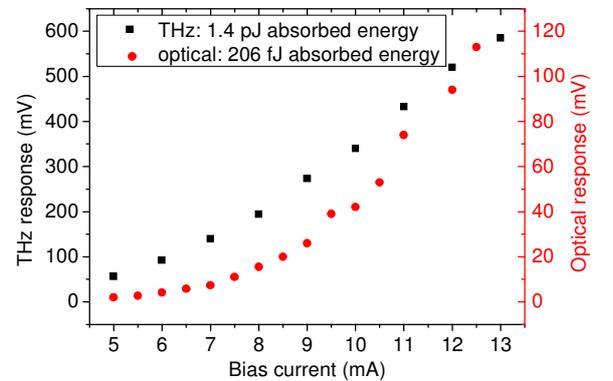

FIG. 6. (Color online) Transient magnitudes of the 30 nm thick YBCO sample in the THz-regime (squares) and optical frequency range (circles). The optical response shows a non-linear dependence in contrast to the linear dependence in the THz frequency range.

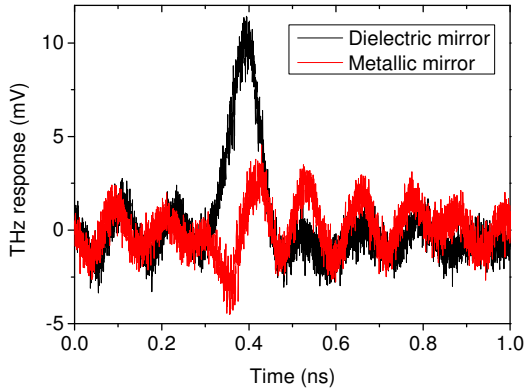

FIG. 7. (Color online) THz detector response of the YBCO sample at zero bias when replacing the metallic mirror with the dielectric STO mirror. The sign of the transient changes due to the change of the phase of the THz electric field.

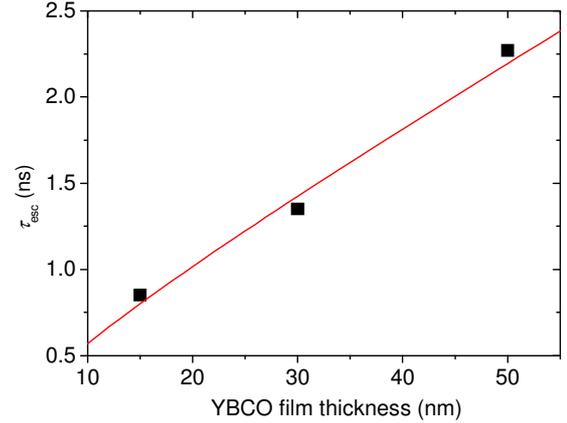

FIG. 8. (Color online) Phonon escape time of the YBCO thin-film samples for optical radiation ($\lambda = 800$ nm). The solid line shows the linear fit of $\tau_{esc}$ in dependence of the effective film thickness.

The two excitation regimes differ with respect to the ratio of the photon energy and the superconducting energy gap in YBCO. According to Ref. 24 the smaller energy gap of the two-gap superconducting YBCO is 6.3 meV at 0 K ($0.8\,k_B T_c$). It weakly depends on temperature and is still 5.5 meV at $0.9 T_c$ (Ref. 24).

In the optical frequency range at $\lambda = 800$ nm the photon energy of 1.55 eV is well above the energy gap and suffices for destroying cooper pairs.

Contrary, in the THz frequency range the photon energy is much smaller than the energy gap. The major energy of CSR is emitted in the frequency range below 1 THz. At this frequency the photon energy of 4.1 meV is below the energy gap for all operation conditions.

The bolometric 2T-model fairly well explains the response of our microbridges to optical-light pulses. A clear indication of the bolometric nature of the optical response is the dependence of the phonon escape time $\tau_{esc}$ on the microbridge thickness. The escape time was determined by the exponential fit of the slow component of the response transient for three different film thicknesses (15, 30 and 50 nm). According to Ref. 25 $\tau_{esc}$ scales with the effective thickness $d_{eff} = d\,[1+d_2/(2d+d_2)]$ which accounts also for the phonons first escaping from the YBCO layer into the PBCO protection layer with the thickness $d_2$ and after reflection from the surface going back through the total thickness $d+d_2$ into the substrate. Fig. 8 shows the fit of the measured escape times to this model. The scaling parameter 0.036 ns/nm, which is the escape time per 1 nm of the layer thickness, was found to be in good agreement with data reported for thicker YBCO films[25, 26].

We used typical parameters of the YBCO bridges and the pulse duration of 10 ps to model the response to CSR pulses in the framework of the 2T-model. Adjusting only the absorbed energy it was not possible to reproduce neither the magnitude nor the shape of the response transients obtained in the THz regime.

One possible explanation could be that the microbridge is only partly heated leading to a local bolometric response where the relaxation is dominated by electron diffusion without any significant contribution from phonon cooling. Using the known readout transfer function and the measured transient amplitudes, we estimated the length of the microbridge fraction, which has to be driven to the normal state by the THz CSR. The required length is approximately 100 nm for the largest energy of CSR pulses. With the diffusion coefficient $D = 0.8$ cm$^2$/s (Ref. 27) the cooling time via electron diffusion will be a few times larger than the electron-phonon interaction time at this temperature. Therefore, the slow bolometric component should be present even if the response is provided by local heating. Thus, we rule out the possibility of the local bolometric response.

The zero-bias response to THz CSR pulses and the polarity change, which follows the phase of the electric field, evidences that the response in the THz regime is rather controlled by the electric field of the CSR pulse. To get insight into the field structure of the CSR pulses, we computed the electric field transient, which is emitted by just one electron with an energy of 630 MeV (corresponding orbit radius of 1.528 m) when it passes the bending magnet of MLS. The observation point is placed in the orbit plane and since in this case and for low frequencies ¾ of the spectral power density is polarized parallel to that plane we consider only the horizontal component of the electric field. The electric field is then filtered with a bandpass from 0.1 to 1 THz (box-car filter) representing the antenna lower frequency cut-off and the upper frequency boundary of CSR, respectively. In fig. 9 the time evolution of the electric field is shown. The main negative peak is followed by a smaller positive one. The field relevant for the transients in fig. 7 is not the field from only one electron but the convolution of fields from all electrons in a bunch. The convolution leads to a smearing of the pulse shape and a relative reduction of the positive peak. Nevertheless, the overall shape of the bunch field remains close to that from a single electron. The zero bias response shown in fig. 7 falls in with this expected shape that supports the idea of the detection mechanism sensitive to the electric field of the CSR.

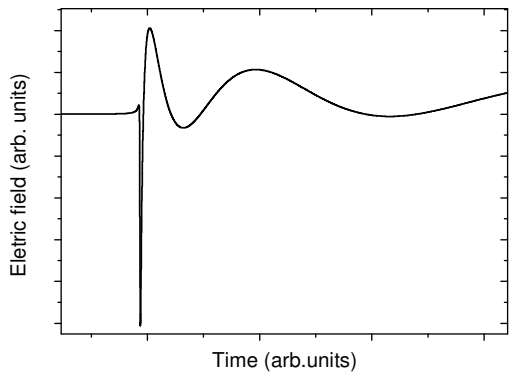

FIG. 9. Simulated electric THz-field transient of one electron filtered by a bandpass from 0.1 to 1 THz for MLS.

We now estimate the current which is induced in the microbridge by the electric field of the CSR pulse. Equating the coupled pulse energy to the Joule energy, which is dissipated within the pulse duration in the microbridge with the normal resistance, we obtained for the coupled energy of 1.4 pJ the current amplitude a few times larger than the bias current. This current is definitely larger than the current at which the vortex-flow instability sets on[28]. In the current-voltage characteristic (fig. 4) the onset of the vortex-flow instability is seen as a discontinuous jump in the voltage. Before the onset of the instability the dc resistance of the microbridge $R$ amounts to approximately $0.1\,\Omega$. Assuming that this resistance is due to the movement of vortices, we estimate the corresponding vortex velocity as $2\pi d \ln(w/2d)R/(\mu_0 l) \approx 3\cdot10^4$ m/s where $l$ and $w$ is the length and width of the microbridge. We also estimate the critical vortex velocity with the Larkin-Ovchinnikov (LO) theory[29] as $v_C \approx (D/\tau_{in})^{1/2}$. Here $\tau_{in}$ is the inelastic scattering time, which we associate with the electron-phonon interaction time. With our parameters we find $v_C = 3\cdot10^3$ m/s. Taking into account the increase of the critical velocity with the decrease of the magnetic field, we find values close to the vortex velocity right before the onset. We further estimate the magnitude of the voltage transient $\approx \Phi_0\, v_C/w$ ($\Phi_0$ is the flux quantum) that corresponds to a single vortex crossing the microbridge with the critical velocity and obtain the amplitude of $2.5\,\mu\text{V}$ on the microbridge. This is still much lower than the experimentally observed transient magnitude at zero bias ($70\,\mu\text{V}$) which we calculated taking into account the known readout transfer function and the measured transient amplitude of 3 mV at zero bias with the metallic mirror (see fig. 7). We note, however, that the actual vortex velocity may be higher than we have estimated. It has been shown that vortices in the LO-limit change their shape and can not be any more described as conventional equilibrium Abricosov vortices. Additionally, several vortices in a row may cross the microbridge[30]. Vortex velocity by two orders of magnitude larger than $v_C$ was found by the numerical solution of the time-dependent Ginzburg-Landau (TDGL) equations[31] for the dimensionless parameters close to those of our microbridges. On the other hand, even this velocity is not high enough to allow the vortex crossing the microbridge completely within the pulse duration.

Anyway, we anticipate that the THz current pulse causes a rearrangement of the vortex lattice into a vortex line[30] which creates the channel with reduced superconducting order parameter. Vortices (of the LO type or kinematic) cross the microbridge via this channel giving the voltage transient. In the presence of the bias current healing of this channel lasts longer than $\tau_{in}$. The actual healing time increases with the bias current[32]. Therefore, the response magnitude in the presence of the bias current is defined by this current and the reduction of the order parameter in the channel, which was created by the THz current pulse. Such a scenario explains qualitatively our experimental data. However, we would like to stress that the proposed model implies the solution of TDGL equations in the limiting case of weak electron-electron interaction. Whether this holds for YBCO is not clear. Analysing TDGL equations with strong electron-electron interaction may help to quantitatively describe the vortex-assisted photoresponse of YBCO microbridges to THz synchrotron radiation.

## VII. CONCLUSION

We have shown that the photoresponse of YBCO microbridges to short CSR pulses in the THz regime drastically differs from the conventional bolometric response of the very same samples to short optical pulses. We have shown that the conventional two-temperature bolometric model fails to describe the response in the THz regime. Finally, we proposed the scenario of vortex-assisted photoresponse that qualitatively explains our experimental results. In brief, the electric field of the CSR pulse results in a current pulse through the microbridge that creates a channel for vortices to cross the microbridge. The response magnitude is defined by the bias current and the reduction of the order parameter imposed by the THz field.

## ACKNOWLEDGMENTS

This work was funded by the German Federal Ministry of Education and Research (Grant No. 05K2010). Authors thank D. Vodolazov for stimulating discussions.